\documentclass[
    ,final            
  ]
  {aipproc}

\layoutstyle{8x11single}

\newcommand{\nn}{\nonumber}
\newcommand{\be}{\begin{equation}}
\newcommand{\ee}{\end{equation}}
\newcommand{\ba}{\begin{eqnarray}}
\newcommand{\ea}{\end{eqnarray}}
\newcommand{\ci}[1]{\cite{#1}}

\def\vb0{{\bf b}_0}

\def\mev{\,{\rm MeV}}
\def\gev{\,{\rm GeV}}

\newcommand{\gsim}{\raisebox{-4pt}{$\,\stackrel{\textstyle
                                                         >}{\sim}\,$}}

\newcommand{\req}[1]{(\ref{#1})}
\def\xb{\bar{x}}

\def\={\,=\,}

\begin{document}

\title{GPDs, the structure of the proton and wide-angle Compton scattering}


\author{P.\ Kroll}{
address={Fachbereich Physik, Universit\"at Wuppertal, D-42097 Wuppertal, Germany}
}

\begin{abstract}
Results from a recent analysis of the zero-skewness generalized parton
distributions (GPDs) for valence quarks are reviewed. The analysis
bases on a physically motivated parameterization of the GPDs with a
few free parameters adjusted to the nucleon form factor data. 
Results for the GPDs are presented and a number of applications such
as moments, Ji's sum rule or their impact parameter representation are
discussed. The $1/x$ moments, in particular, form the soft physics input
to Compton scattering off protons within the handbag approach. The 
Compton cross section evaluated from this information is found to be
in good agreement with experiment.\\
Talk presented at the workshop on the Shape of Hadrons, Athens (2006)
 \end{abstract}

\maketitle

\section{Introduction}
In recent years we have learned how to deal with hard exclusive
reactions within  QCD. In analogy to hard inclusive reactions the
process amplitudes factorize in partonic subprocess amplitudes, 
calculable in perturbation theory, and in GPDs which parameterize soft
hadronic matrix elements. In some cases rigorous proofs of
factorization exist. For other processes factorization is shown to
hold in certain limits, under certain assumptions or is just a 
hypothesis. This so-called handbag mechanism applies for instance to
Compton scattering in kinematical regions where either the virtuality
of the incoming photon, $Q^2$, is large while the squared invariant
moment transfer, $t$, from the initial to thte final proton is small
(deeply virtual region) or $-t$ and $-u$ are large but $Q^2$ is small
(wide-angle region). 

The GPDs which are defined by Fourier transforms of
bilocal proton matrix elements of quark field operators, 
describe the emission and reabsorption of partons by the proton. For 
equal helicities of the emitted and reabsorbed parton the structure of
the nucleon is described by four GPDs, termed $H$, $\widetilde{H}$,$E$ 
and $\widetilde{E}$, for each quark flavor and for the gluons. The GPDs
are functions of three variables: the momentum transfer, a momentum 
fraction $x$ and the skewness $\xi$. The latter two variables are 
related to the individual momentum fractions the emitted and
reabsorbed partons carry by 
\be
x_1\= \frac{x+\xi}{1+\xi}\,, \qquad   x_2\= \frac{x-\xi}{1-\xi}\,.
\ee
The GPDs are subject to evolution and, hence, depend on the 
factorization scale $\mu$. They satisfy the reduction formulas
\be
H^q(x,\xi=0,t=0,\mu)\= q(x,\mu)\,, \qquad  
                        \widetilde{H}^q(x,\xi=0,t=0,\mu)\= \Delta q(x,\mu)\,,  
\ee
i.e.\ in the forward limit of zero momentum transfer and zero skewness, 
$H$ and $\widetilde{H}$ reduce to the usual unpolarized and polarized 
parton distributions (PDFs), respectively. The forward limits of $E$
and $\widetilde{E}$ are not accessible in deep inelastic
electron-proton scattering and are therefore unknown as yet. Another 
property of the GPDs is the polynomiality which comes about as a 
consequence of Lorentz covariance 
\be
\int^1_{-1}\, dx\, x^{m-1}\, H^q(x,\xi,t,\mu) \= \sum^{[m/2]}_{i=0}\,
h^q_{m,i}(t)\,\xi^i\,,
\label{pol}
\ee 
where $[m/2]$ denotes the largest integer smaller than or equal to
$m/2$. Eq.\ \req{pol} holds analogously for the other GPDs and, for
$m=1$, implies sum rules for the form factors of the nucleon, e.g.\
\be
F^q_1(t) \equiv h^q_{1,0}(t) \= \int^1_{-1}\, dx H^q(x,\xi,t,\mu)\,,
\label{sumrule}
\ee
which represents the contribution of quarks of flavor $q$ to the Dirac
form factor of the proton. Reinterpreting as usual a parton carrying a 
negative momentum fraction $x$ as an antiparton with a positive $x$ 
($H^{\bar{q}}(x)=-H^q(-x)$), one becomes aware that only the
difference of the contributions from quarks and antiquarks of given 
flavor contribute to the sum rules. Introducing the combination
\be
H^q_v(x,\xi,t,\mu) \=   H^q(x,\xi,t,\mu) -  H^{\bar{q}}(x,\xi,t,\mu)\,,
\ee
for positive $x$ which, in the forward limit, reduces to the usual
valence quark density $q_v(x)=q(x)-\bar{q}(x)$, one finds for the
Dirac form factor the representation
\be
F_1^{p(n)}(t)\= e_{u(d)} \int_0^1\, dx\, H_v^u(x,\xi,t,\mu) + e_{d(u)}
\int_0^1\, dx\, H_v^d(x,\xi,t,\mu)\,.
\label{pn-sr}
\ee     
Here, $e_q$ is the charge of the quark $q$ in units of the positron
charge. Possible contributions from other flavors, $s-\bar{s}$ or 
$c -\bar{c}$, are neglected in the sum rule \req{pn-sr}. They are
likely small: in the forward limit one has $s(x) \simeq \bar{s}(x)$ 
\ci{CTEQ} and the strange form factors of the nucleon are seemingly
small \ci{happex}. This simplification of the sum rule \req{pn-sr} 
does not imply that the nucleon is assumed to consist solely of 
valence quarks. Sea quarks are there but the virtual photon that 
probes the quark content of the nucleon, sees only the differences 
between quark and antiquark distributions which are likely small 
for strange and charm quarks.

A representation analogue to \req{pn-sr} also holds for the 
Pauli form factor with $H$ being replaced by $E$. The isovector 
axial-vector form factor, on the other hand, satisfies the sum rule 
($\widetilde{H}^{\bar{q}}(x)=\widetilde{H}^{q}(-x)$)
\be
F_A(t) \= \int_0^1 dx\, \left\{\Big[\widetilde{H}_v^u(x,\xi,t,\mu) -
                                  \widetilde{H}_v^d(x,\xi,t,\mu)\Big]
                   + 2 \, \Big[\widetilde{H}^{\bar{u}}(x,\xi,t,\mu) -
                       \widetilde{H}^{\bar{d}}(x,\xi,t,\mu)\Big]\right\}\,,
\label{axial-sr}
\ee
At least for small $-t$ the magnitude of the second integral in Eq.\ 
\req{axial-sr} reflects the size of the flavor non-singlet combination 
$\Delta \bar{u}(x) - \Delta \bar{d}(x)$ of forward densities. This 
difference is poorly known, and at present there is no experimental 
evidence that it might be large \ci{HERMES}. For instance, in the
analysis of the polarized PDFs performed by Bl\"umlein and 
B\"ottcher~Ref.\ \ci{BB} it is zero. Hence, in a perhaps crude 
approximation the second term in Eq.\ \req{axial-sr} may be neglected. 
\section{Extracting the zero-skewness GPDs}
Since the GPDs cannot be calculated from QCD with a sufficient degree of
accuracy at present we have either to rely on models or to extract
them from experiment as it has been done for the PDFs, see for
instance Refs.\ \ci{CTEQ,BB}. 
First attempts to extract the GPDs phenomenologically have been
carried through in Refs.\ \ci{DFJK4} and \ci{guidal}. The main idea is to 
exploit the sum rules \req{pn-sr} and \req{axial-sr} and to determine
the GPDs from the nucleon form factor data, $F_1$ and $F_2$ for proton 
and neutron as well as $F_A$ (for references to the data see Ref.\
\ci{DFJK4}). Since the sum rules represent only the first moments of
the GPDs this task is an ill-posed problem in a strict mathematical sense. 
Infinitely many moments are needed to deduce the integrand, i.e.\ the
GPDs, unambigously from an integral. However, from phenomenological 
experience with particle physics one expects the GPDs to be smooth
functions and a small number of moments will likely suffice to fix the
GPDs. The extreme - and at present the only feasible - point of view
is to assume that the lowest moment of a GPD alone is sufficient to fix it 
\ci{DFJK4,guidal}. Indeed, using recent results on PDFs~\ci{CTEQ,BB}
and form factor data in combination with physically motivated 
parameterizations of the GPDs, one can carry through such an analysis. 
Needless to say that this method while phenomenologically succesful 
as will be discussed below, does not lead to a unique result. Other 
parameterizations which may imply different physics, cannot be
excluded at the present stage of the art. 

The sum rules \req{pn-sr} and \req{axial-sr} hold for all $\xi$ but
guessing a plausible parameterization of a function of three variables
is nearly hopeless task. In order to simplify matters one may
choose the special value $\xi=0$ for which the emitted and reabsorbed 
partons carry the same momentum fractions. This choice  
has many advantages. One exclusively works in the so-called DGLAP
region where $\xi\leq x$. In this region parton ideas apply, the Fourier
transform of the GPDs with respect to the momentum transfer 
$\Delta$ ($\Delta^2=-t$) has a density interpretation in the
impact parameter plane~\ci{burk} and, as shown in Refs.\ 
\ci{DFJK1,rad98}, wide-angle exclusive processes are controlled by 
generalized form factors which represent $1/x$ moments of zero-skewness 
GPDs. The parameterization that is exploited in Ref.\ \ci{DFJK4}
combines the usual PDFs with an exponential $t$ dependence (the 
arguments $\xi=0$ and $\mu$ are omitted in the following for convenience)
\be
H_v^q(x,t) \= q_v(x) \exp[tf_q(x)]\,,
\label{ansatz}
\ee
where the profile function reads 
\be
f_q(x) \= \big[\alpha'\, \log(1/x) + B_q\big](1-x)^{n+1} + A_q x (1-x)^n\,.
\label{profile}
\ee 
This ansatz is motivated by the expected Regge behaviour at low $-t$
and low $x$ \ci{arbarbanel} (where $\alpha'$ is the Regge slope for
which the value $0.9\,\gev^2$ is imposed at low factorization scales). 
At large $-t$ and large
$x$, on the other hand, one expects a behaviour like $f_q \sim 1-x$
from overlaps of light-cone wavefunctions~\ci{DFJK1,DFJK3}. The ansatz 
\req{ansatz}, \req{profile} interpolates between the two limits 
smoothly~\footnote{
The parameter $B_q$ is not needed if $\alpha'$ is freed. A fit to the
data of about the same quality and with practically the same results
for the GPDs is obtained for $\alpha^\prime\simeq 1.4$.} 
and allows for a  stronger suppression of $f_q$ in the limit $x\to 1$.  
It matches the following criteria for a good parameterization:
simplicity, consistency with theoretical and phenomenological
constraints, stability with respect to variations of PDFs and
stability under evolution (scale dependence of the GPDs can be absorbed 
into parameters).

\begin{figure}
\includegraphics[width=.30\textwidth, height=.33\textwidth,
  bb=77 448 399 786,clip=true] {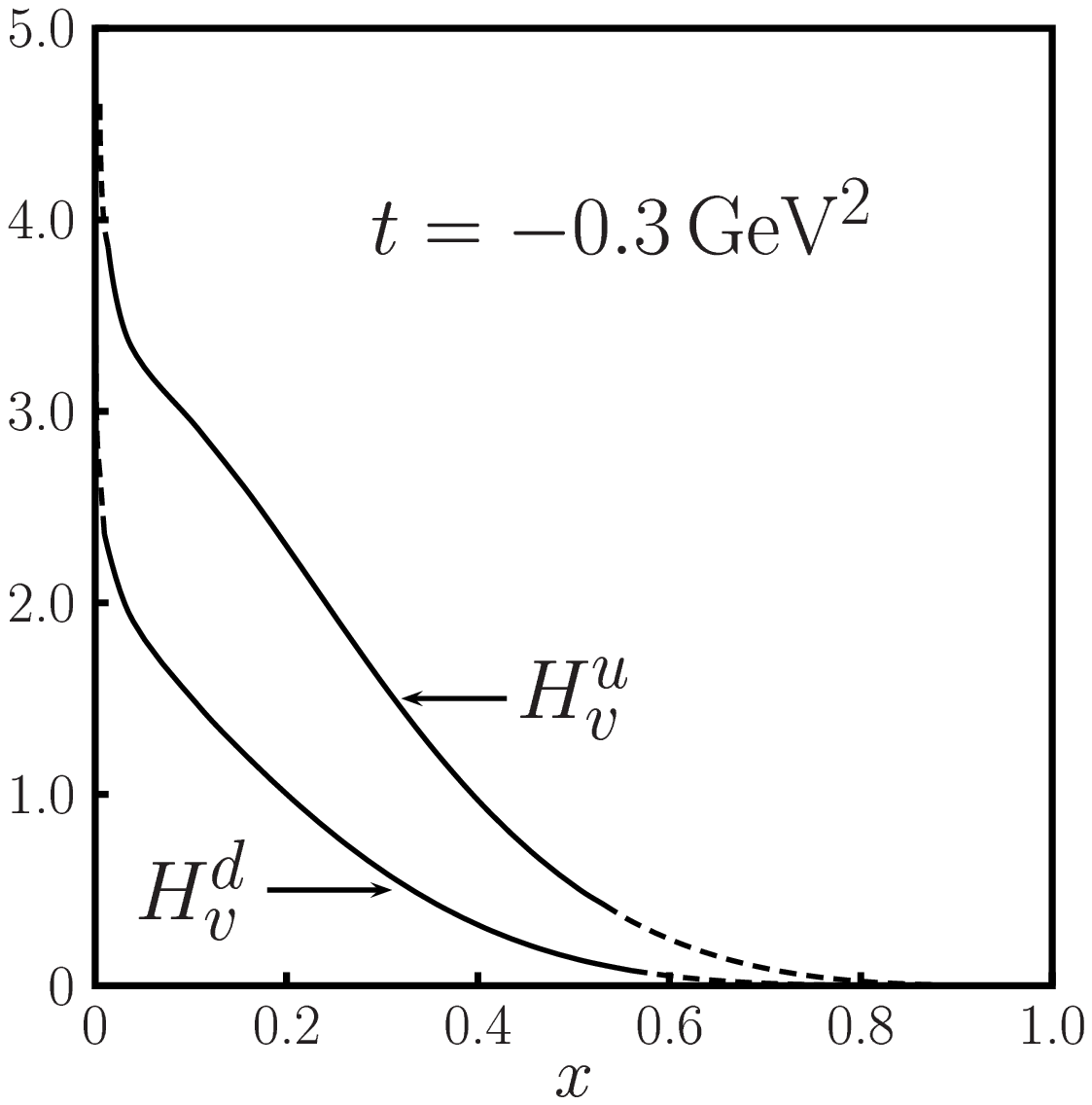}
\hspace{1em} 
\includegraphics[width=.30\textwidth, height=.33\textwidth,
  bb=76 294 400 628,clip=true] {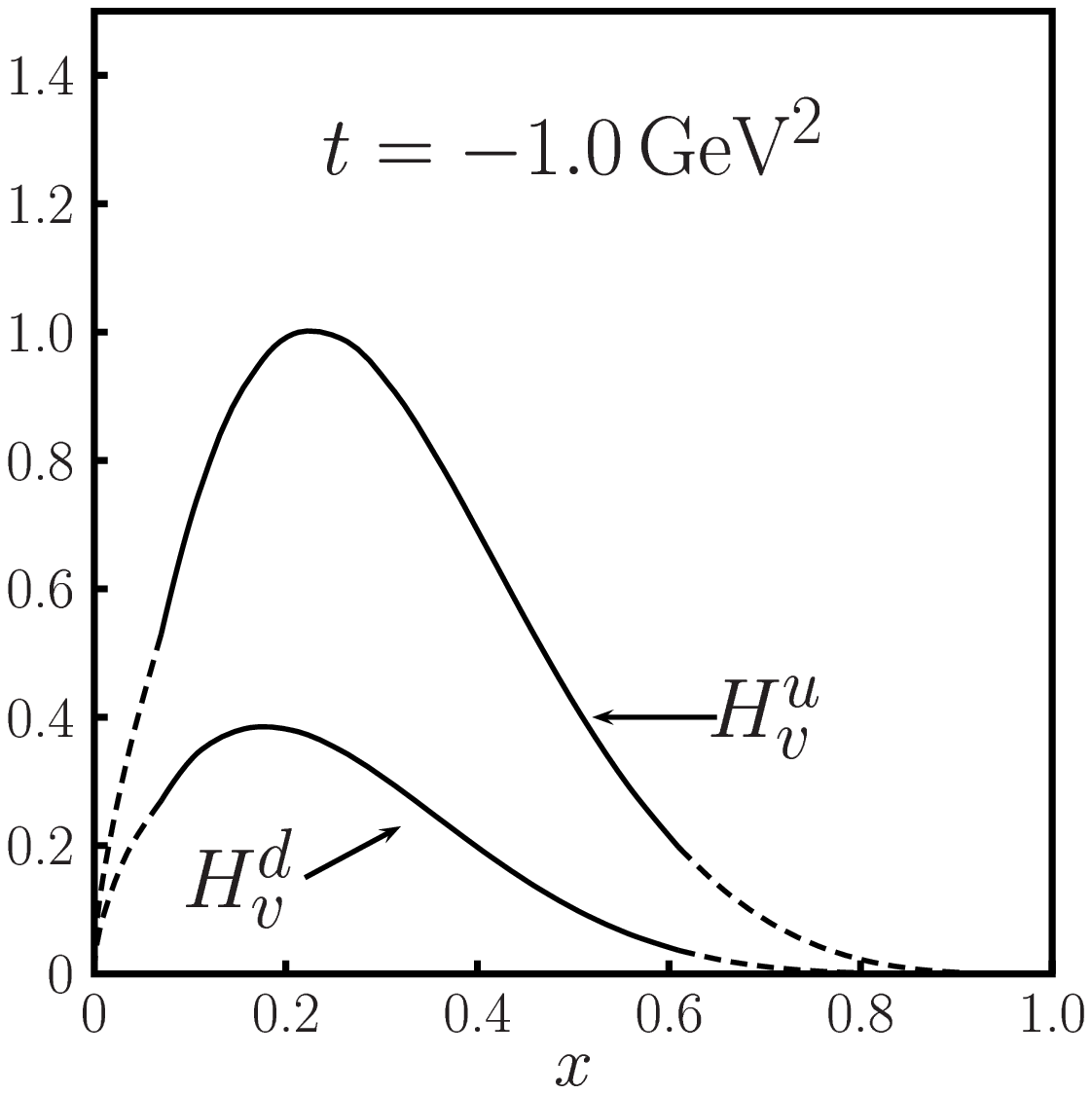}
\hspace{1em} 
\includegraphics[width=.30\textwidth, height=.33\textwidth,
  bb=66 479 400 826,clip=true] {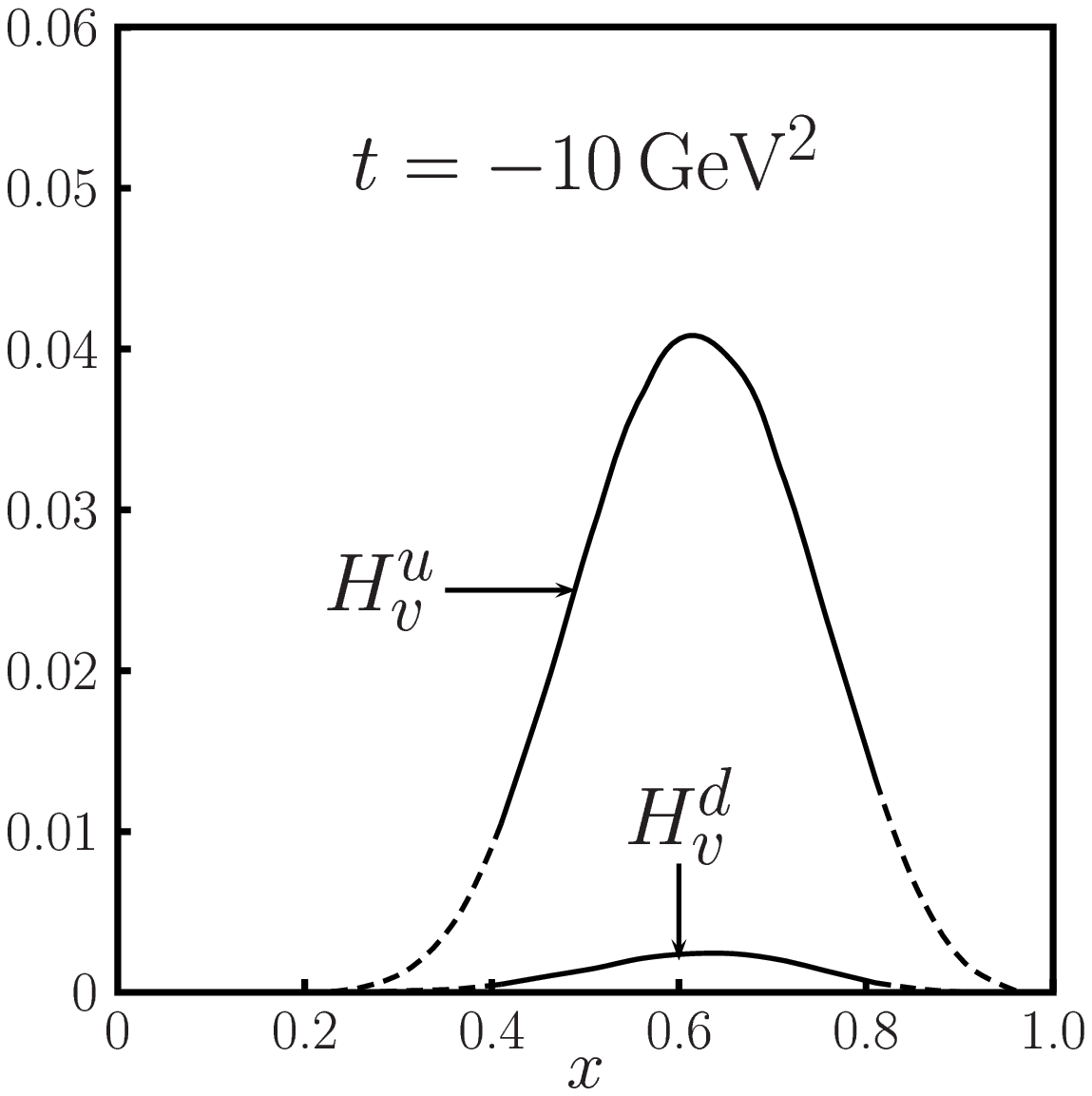}
\vspace*{-1.6em}
\caption{\label{fig:Hgpd} Results for $H_v^q(x,t)$ at the scale 
$\mu=2\,\gev$ and for $n=2$ obtained in \cite{DFJK4}. In each of the
  regions indicated by dashed lines, $5\%$ of the total value of the 
sum rule \req{pn-sr} are accumulated.}
\label{fig:gpdH}
\end{figure}

Using the CTEQ PDFs \ci{CTEQ}, one obtains a reasonable fit of the
ansatz \req{ansatz}, \req{profile} to the data on the Dirac form
factor which range from $-t=0$ to $\simeq 30\, \gev^2$, with the
parameters
\ba
B_u&=&B_d\=(0.59\pm 0.03)\,\gev^{-2}\,, \nn\\
 A_u&=&(1.22\pm 0.020)\,\gev^{-2}\,, \quad A_d= (2.59\pm 0.29)\,\gev^{-2}\,, 
\ea
quoted for the case $n=2$ and at a scale of $\mu=2\,\gev$. In 
Fig.\ \ref{fig:gpdH} the results for $H^q_v$ are shown at three values
of $t$. While at small $-t$ the behaviour of the GPD still reflects
that of the parton densities it exhibits a pronounced maximum at
larger values of $-t$. The maximum becomes more pronounced with
increasing $-t$ and its position moves towards $x=1$. In other words
only a limited range of large $x$ contributes to the form factor
substantially.
The quality of the fit is very similar in both the cases, $n=1$ and 2;
the results for the GPDs and related quantities agree well with
each other. Substantial differences between the two results only occur
at very low and very large values of $x$, i.e.\ in the regions which are
nearly insensitive to the present form factor data. It is the physical
interpretation of the results which favours the fit with $n=2$. Indeed
the average distance between the struck quark and the cluster of spectators
becomes unphysical large for $x\to 1$ in the case $n=1$; it grows like
$\sim 1/(1-x)$ while, for $n=2$, it tends towards a constant value of
about 0.5 fm \ci{DFJK4}. 

\begin{figure}[t] 
\includegraphics[height=.39\textwidth,
 bb = 123 339 456 673,clip=true]{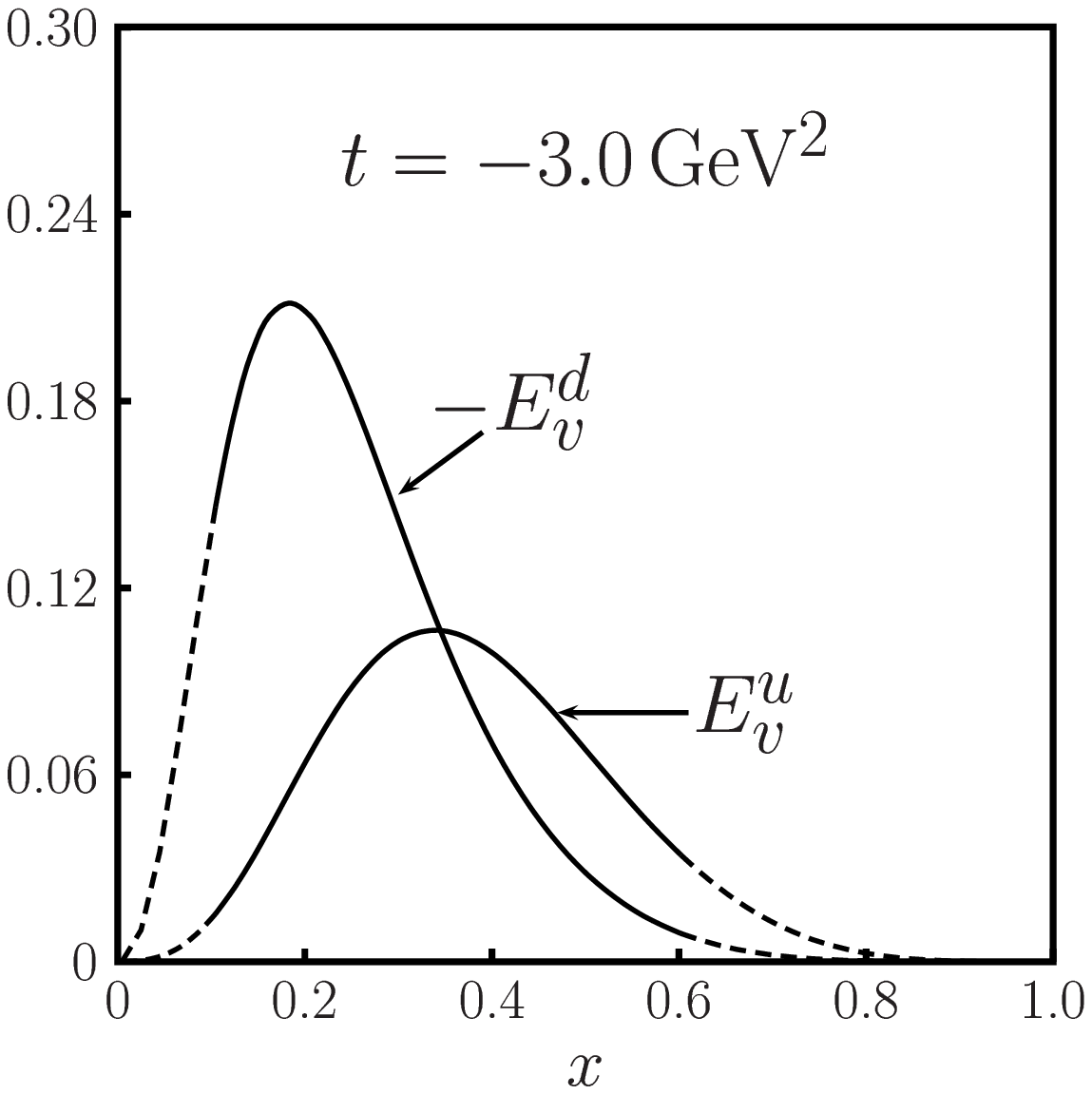}\hspace*{0.5em}
\includegraphics[width =.39\textwidth,
 bb = 104 363 436 695,clip=true]{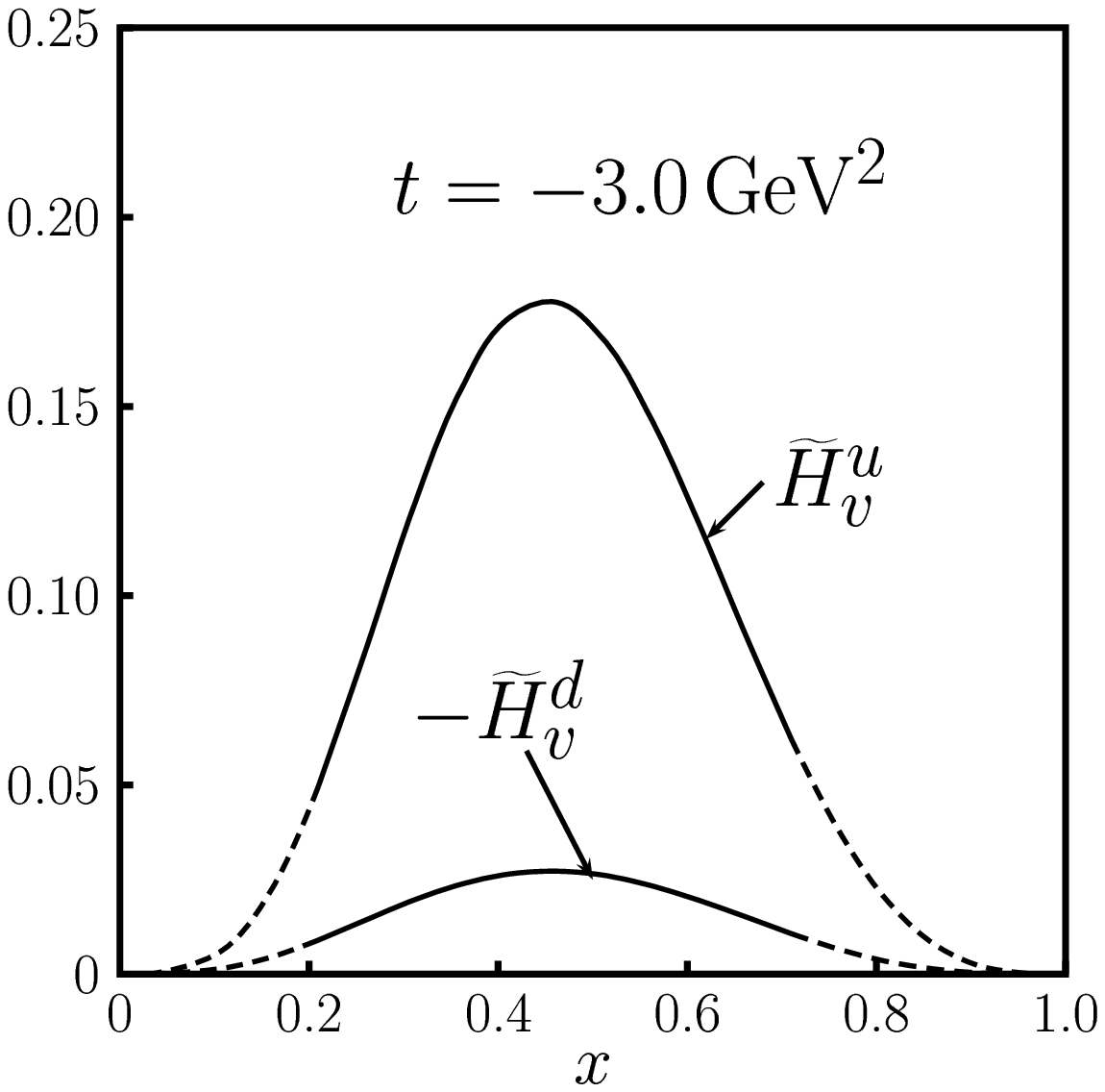}
\vspace*{-1.6em}
\caption{\label{fig:moment} The GPDs $E^{u(d)}_v$ (left) and 
 $\widetilde{H}_v^{u(d)}$ (right) at the scale $\mu=2\,\gev$.}
\end{figure}

The analysis of the axial and Pauli form factors, with
parameterizations analogue to Eqs.\ \req{ansatz}, \req{profile},
provides the GPDs $\widetilde{H}$ and $E$. They behave similar to $H$, 
see Fig.\ \ref{fig:moment}. Noteworthy differences are the opposite
signs of $\widetilde{H}^u_v$ ($E^u_v$) and $\widetilde{H}^d_v$ ($E^d_v$) 
and the relative magnitude of $E^u_v$ and $E^d_v$. At intermediate
values of $-t$ the GPD $-E^d_v$ develops a maximum which is more
pronounced and located at significantly smaller $x$ than in $E^u_v$. 
At larger values of $-t$  $|E^d_v|$ becomes gradually smaller than
$E^u_v$. For $H^q_v$
and $\widetilde{H}^q_v$, on the other hand, the $d$-quark contributions
are substantially smaller in magnitude than the $u$-quark ones, see
Fig.~\ref{fig:gpdH}. Since there is no data available for the
pseudoscalar form factor of the nucleon the GPD $\widetilde{E}$ cannot
be determined this way. 
\begin{figure}[ht]
\includegraphics[height=.39\textwidth,
 bb = 104 357 449 698,clip=true]{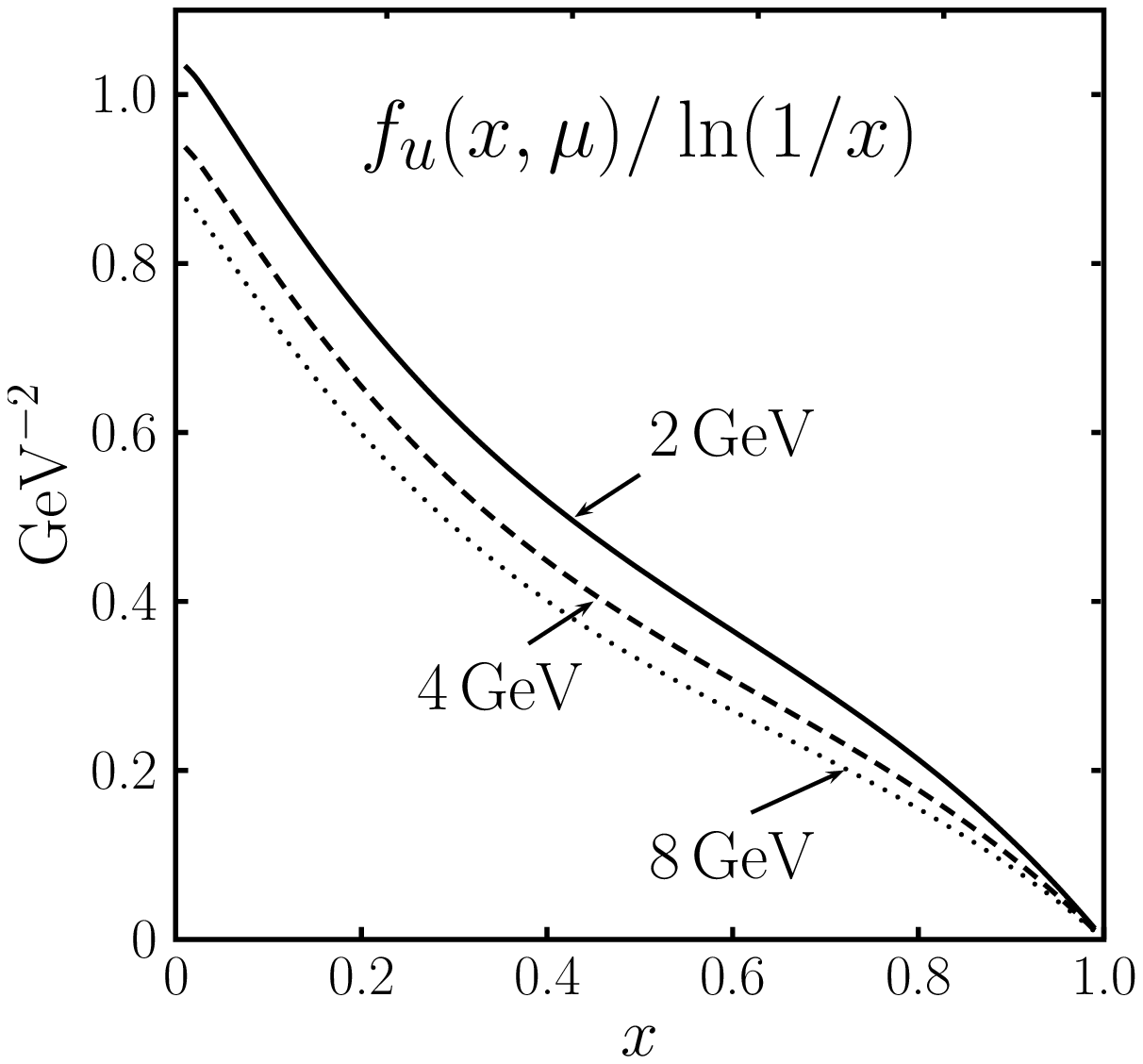}\hspace*{0.5em} 
\vspace*{-1.6em}
\includegraphics[height=.38\textwidth,bb = 67 474 399 817,clip=true]
{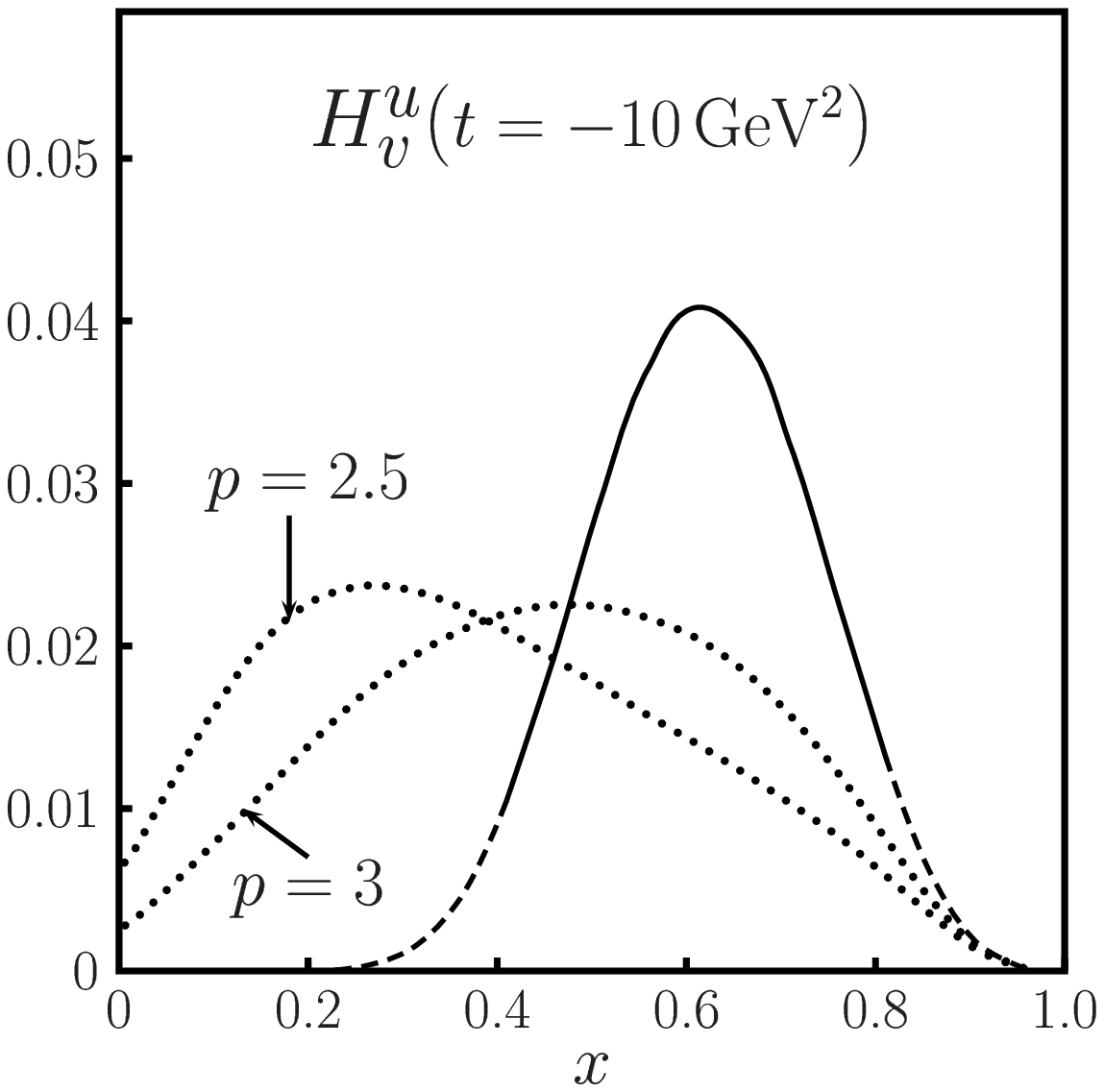} 
\caption{\label{fig:power} The scale dependence of the profile
  function for $u$ quarks (left) and the $H_v^u$ at $t=-10\,\gev^2$
  obtained with the parameterization \req{ansatz-power} for the powers
  $p=2.5$ and 3 (right). These results are compared to the GPD
  obtained with the exponential profile function \req{ansatz}, \req{profile}.}
\end{figure}

The scale dependence of the profile function for $u$ quarks is shown in
Fig.\ \ref{fig:power}. Combined with that of the PDF it provides the
scale dependence of $H^u$. Analogue results are found for $d$ quarks
and for the other GPDs. The profile functions decrease with increasing 
$\mu$, i.e., the nucleon becomes more compact in the impact parameter 
plane at larger scales.

Up to now only the sum rules \req{pn-sr}, \req{axial-sr} have been
utilized in the GPD analysis. As mentioned above, in this situation 
parameterizations of the GPDs are required with the consequence of
non-unique results. An alternative ansatz~\ci{DFJK4} is for instance 
\be
H_v^q(x,t) \= q_v(x)\Big[1-t f_q(x)/p\Big]^{-p}\,.
\label{ansatz-power}
\ee
Although reasonable fits to the form factors are obtained with 
\req{ansatz-power} for $p\gsim 2.5$ it is physically less
appealing than the parameterization \req{ansatz}: the combination of 
Regge behaviour at small $x$ and $t$ with the dynamics of the Feynman 
mechanism is lost. The resulting GPDs have a broader shape and 
$H(x=0,t)$ remains finite, see Fig.\ \ref{fig:power}. Thus, small $x$ 
also contribute to the high-$t$ form factors for this parameterization. 
The parameterization \req{ansatz-power} is not stable under DGLAP evolution.

\section{Applications}
\begin{figure}
\includegraphics[height=.39\textwidth,
 bb = 48 296 444 655,clip=true]{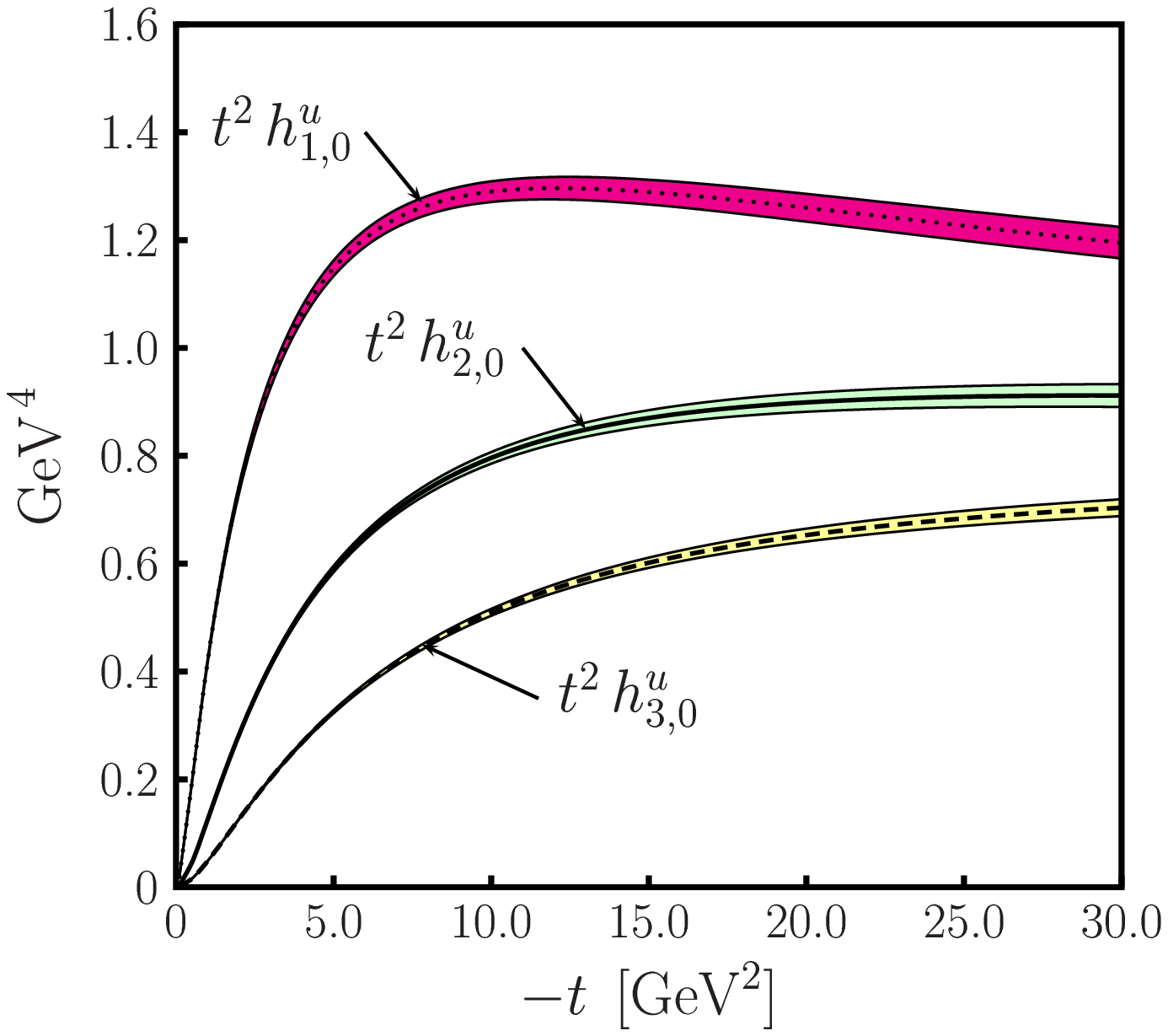}\hspace*{0.5em}
\includegraphics[width =.39\textwidth,
 bb = 103 338 447 677,clip=true]{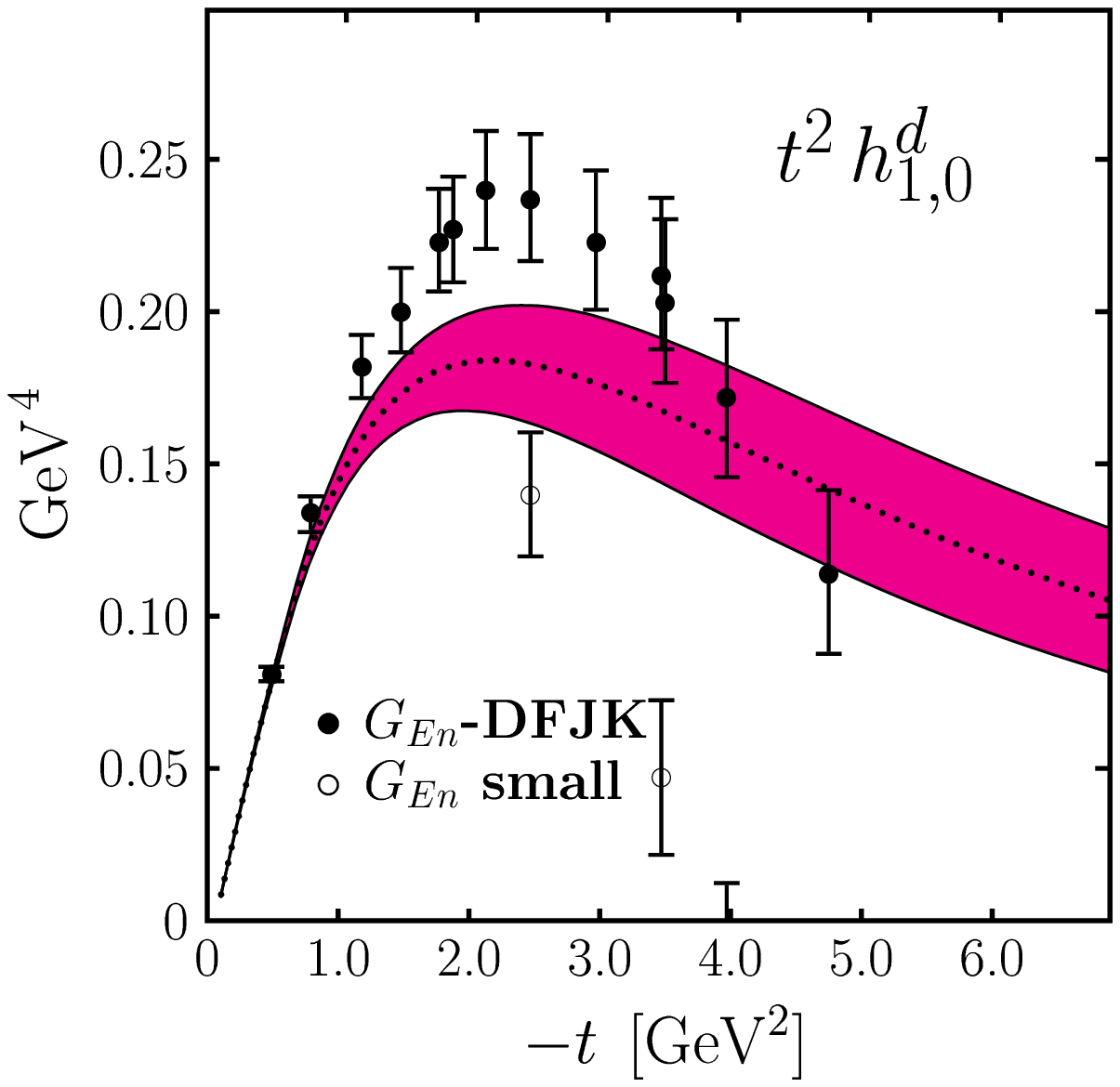}
\vspace*{-1.6em}
\caption{\label{fig:moment2} The first three moments
  of $H_v^u$ (left) and the lowest moment of $H_v^d$ (right) 
  scaled by $t^2$. The shaded bands represent the
  parametric uncertainties \ci{DFJK4}. Data are taken from 
  Ref.\ \ci{CLAS} and from the references quoted in
  \ci{DFJK4}. Results for two different extrapolations of $G_E^n$ are shown.}
\end{figure}

Having the GPDs at disposal one can evaluate various moments, some of
them are displayed in Fig.\ \ref{fig:moment2}. 
An interesting property of the moments is that the $u$ and $d$ quark
contributions decrease with different rates at large $-t$. Since in 
this region the dominant contribution to the form factors comes from a 
narrow region of large $x$ (see Fig.\ \ref{fig:gpdH}) one can use the 
large $x$ approximations (given for the favored case $n=2$)
\be
q_v \sim (1-x)^{\beta_q}\,, \qquad f_q \sim A_q(1-x)^2\,,
\ee
and evaluate the sum rule \req{pn-sr} in the saddle point approximation. 
This leads to
\be
h^q_{1,0} \sim |t|^{-(1+\beta_q)/2}\,, \quad 
       1-x_s\= \left(\frac{2}{\beta_q}A_q|t|\right)^{-1/2}\,,
\label{power}
\ee
where $x_s$ is the position of the saddle point. The latter lies within 
the region where the GPD is large and using the the CTEQ values for 
$\beta_q$ ($\beta_u\simeq 3.4$ and $\beta_d\simeq 5$ \ci{CTEQ}), one 
obtains a fall off slightly faster than $t^{-2}$ for the form factor 
$h^u_{1,0}$  while the $d$-quark form factor falls as $t^{-3}$. 
Strengthened by the charge factor the $u$-quark contribution
consequently dominates the proton's Dirac form factor~\footnote{
This implies the ratio $F_1^n/F_1^p = e_d/e_u$ at large $-t$.}
for $-t$ larger than about $5\,\gev^2$, the $d$-quark contribution
amounts to less than $10\%$. The power behaviour \req{power} bears 
resemblance to the Drell-Yan relation~\ci{DY}. In fact the common
underlying dynamics is the Feynman mechanism which applies in the soft
region where $1-x\sim \Lambda/\sqrt{-t}$ and the virtualities of the
active partons are $\sim \Lambda\sqrt{-t}$ ($\Lambda$ is a typical 
hadronic scale of order $1\,\gev^2$), cf. Eq.\ \req{power}. The 
Drell-Yan relation is, however, an asymptotic result ($x\to 1$, 
$t\to -\infty$) which bases on the assumption of valence Fock state 
dominance, i.e.\ on the absence of sea quarks. The different powers
for $u$ and $d$ quarks signal that the asymptotic region where the 
dimensional counting rules apply, has not yet been reached. 

One may object to the different powers that they are
merely a consequence of the chosen parameterization \req{ansatz},
\req{profile}. However, this is likely not the case. The moments
$h_{1,0}^q$ can directly be extracted from the form factor
data. The experimental results for the $d$-quark moment are
shown in Fig.\ \ref{fig:moment2}. The sharp fall off with increasing
$-t$ is clearly visible. The small deviations between data and the
moment obtained in \ci{DFJK4} are caused by the very precise but still
preliminary CLAS data~\ci{CLAS} on $G_M^n$. These data have been utilized
in the extraction of the moments $h_{1,0}^q$ but not in the analysis
in Ref.\ \ci{DFJK4}. Worst measured of the four form factors is the
electric one of the neutron. Above $1.5\,\gev^2$ no data is available
as yet and two different extrapolations to larger $-t$ are used
in the extraction of the moments $h_{1,0}^q$. Thus, except $G_E^n$
differs from expectation strongly, the rapid decrease of $h_{1,0}^d$ 
seen in Fig.\ \ref{fig:moment2}, seems to be an experimental
fact. Future JLab data on $G_E^n$ will settle this question.  

A combination of the second moments of $H$ and $E$ at
$t=0$ is Ji's sum rule \cite{ji97} which allows for an evaluation of 
the valence quark contribution to the orbital angular momentum the quarks 
inside the proton carry
\be
\langle L_v^q\rangle \= \frac12\,\int_0^1 dx
       \Big[xE_v^q(x,t=0) + x q_v(x)- \Delta q_v(x)\Big]\,.
\ee
The analysis of the three GPDs leads to 
\be
\langle L_v^u \rangle \= -(0.24 - 0.27)\,,\quad \langle L_v^d \rangle
\= 0.15 - 0.19\,,
\ee
for the valence quark contributions to the orbital angular momentum at
a scale of $\mu=2\,\gev$. The signs of these predictions are in accord
with recent results from lattice QCD~\ci{SESAM,qcdsf} but the absolute 
values are somewhat larger. Given the uncertainties in all these
analyses, Ref.\ \ci{DFJK4} and the lattice calculations, one may
ascertain fair agreement. There is also remarkable agreement of the $t$
dependencies of the moments obtained in Ref.\ \ci{DFJK4} and in
lattice QCD studies. 

\begin{figure}[t]
\includegraphics[width =.40\textwidth,
bb = 124 82 314 273,clip=true]{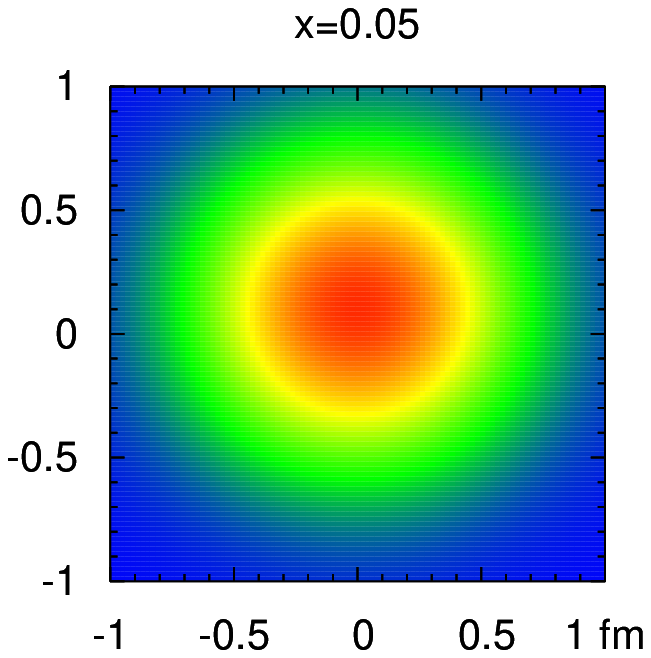} \hspace*{2em} 
\includegraphics[width=0.4\textwidth,bb=124 82 314 273,clip=true]
{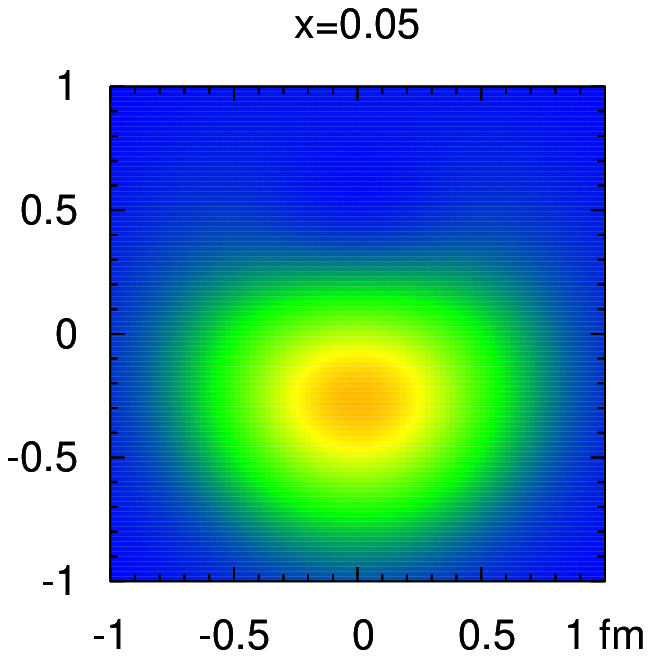}
\vspace*{-1.6em}
\caption{\label{fig:neutron} The $u$ (left) and $d$ (right) quark
  densities in the impact parameter plane at $x=0.005$. The proton is
  polarixed in $X$ direction, see text.}
\end{figure}
While the parton distributions only provide information on the
longitudinal distribution of quarks inside the nucleon, GPDs also give 
access to the transverse position distributions of partons within the
proton. Thus, the Fourier transform of $H$ for instance 
\be
q_v(x,{\bf b}) \= \int \frac{d^2{\Delta}}{(2\pi)^2}\, 
                e^{-i{\bf b}\cdot{ \Delta}}\, H_v^q(x,t=-\Delta^2)\,,
\ee  
gives the probability of finding a valence quark with longitudinal momentum
fraction $x$ and impact parameter ${\bf b}$ as seen in a frame in
which the proton moves rapidly in the $z$ direction. Together with the
analogue Fourier transform of $E_v^q(x,t)$ one can form the
combination ($m_p$ being the mass of the proton)
\be
q_v^X(x,{\bf b}) \= q_v(x,{\bf b}) - \frac{\;b^Y}{m_p}\,
\frac{\partial}{\partial {\bf b}^2}\, e_v^q(x,{\bf b})\,,
\ee
which gives the probability to find an unpolarized valence quark with
momentum fraction $x$ and impact parameter ${\bf b}=(b^X,b^Y)$ in a
proton that moves rapidly along the $Z$ direction and is polarized
in $X$ direction \ci{burk}. As shown in Ref.\ \ci{DFJK4}, for
small $x$ one observes a very broad distribution while at large $x$ 
it becomes more focussed on the center of momentum defined by 
$\sum_i x_i {\bf b}_i=0$ ($\sum_i x_i=1$). In a proton that is
polarized in the $X$ direction the symmetry around the $Z$ axis is
lost and the center of the density is shifted in the $Y$ direction
away from the center of momentum, downward for $d$ quarks and upward 
for $u$ ones. Thus, a polarization of the proton induces a flavor 
segregation in the direction orthogonal to the direction of the 
polarization and the proton momentum, see Fig.\
\ref{fig:neutron}. This effect may be responsible for the Sivers
function \ci{sivers}, $f_{1T}^{\perp }$. One may argue \ci{burkardt06} that the
average Sivers function for quarks of flavor $q$ is related to the
contribution of these quarks to the anomalous magnetic moment by
\be
f_{1T}^{\perp q} \sim - \kappa_q\,,
\ee
where 
\be
\kappa_q \= \int_0^1 d\xb\, E^q_v(\xb,0,0)\,.
\ee
The anomalous moments of the quarks are linear combination of those of
proton and neutron
\be
\kappa_u \= 2\kappa_p + \kappa_n\simeq 1.67\,, \qquad  
\kappa_d \= \kappa_p + 2\kappa_n\simeq -2.03\,.
\ee
This explains the different signs of the Sivers functions fir $u$ and
$d$ quarks and is also the origin of the opposite signs of $E_v^u$ and
$E_v^d$ (see Fig.\ \ref{fig:moment}). The opposite signs of the Sivers 
functions is in agreement with a recent HERMES measurement \ci{HERMES05}.
        
\section{Wide-angle Compton scattering}
As discussed in previous sections the analysis of the GPDs gives
insight in the transverse distribution of quarks inside the proton. 
However, there is more in it. The universality property of the GPDs, 
i.e.\ their process independence, allows to predict other hard 
exclusive reactions once the GPDs have been determined in the analysis 
of a given process. This way QCD acquires a predictive power for hard 
exclusive processes provided factorization holds.

Thus, with the $\xi=0$ GPDs at hand one can predict hard wide-angle 
exclusive reactions like Compton scattering or meson photo- and 
electroproduction. For these reactions it is advantagous to work in a
so-called symmetrical frame where the skewness is zero. It has been 
argued \ci{DFJK1,rad98} that, for large Mandelstam variables (
$s, \;-t, \; -u \gg \Lambda^2$), the amplitudes for these processes 
factorize in a hard partonic subprocess, e.g.\ Compton scattering off 
quarks and in form factors representing $1/x$-moments of zero-skewness 
GPDS (see Fig.\ \ref{fig:neutron}). For Compton scattering off protons 
these form factors read 
\begin{figure}[t]
\includegraphics[width=.47\textwidth,
  bb= 50 107 387 430,clip=true]{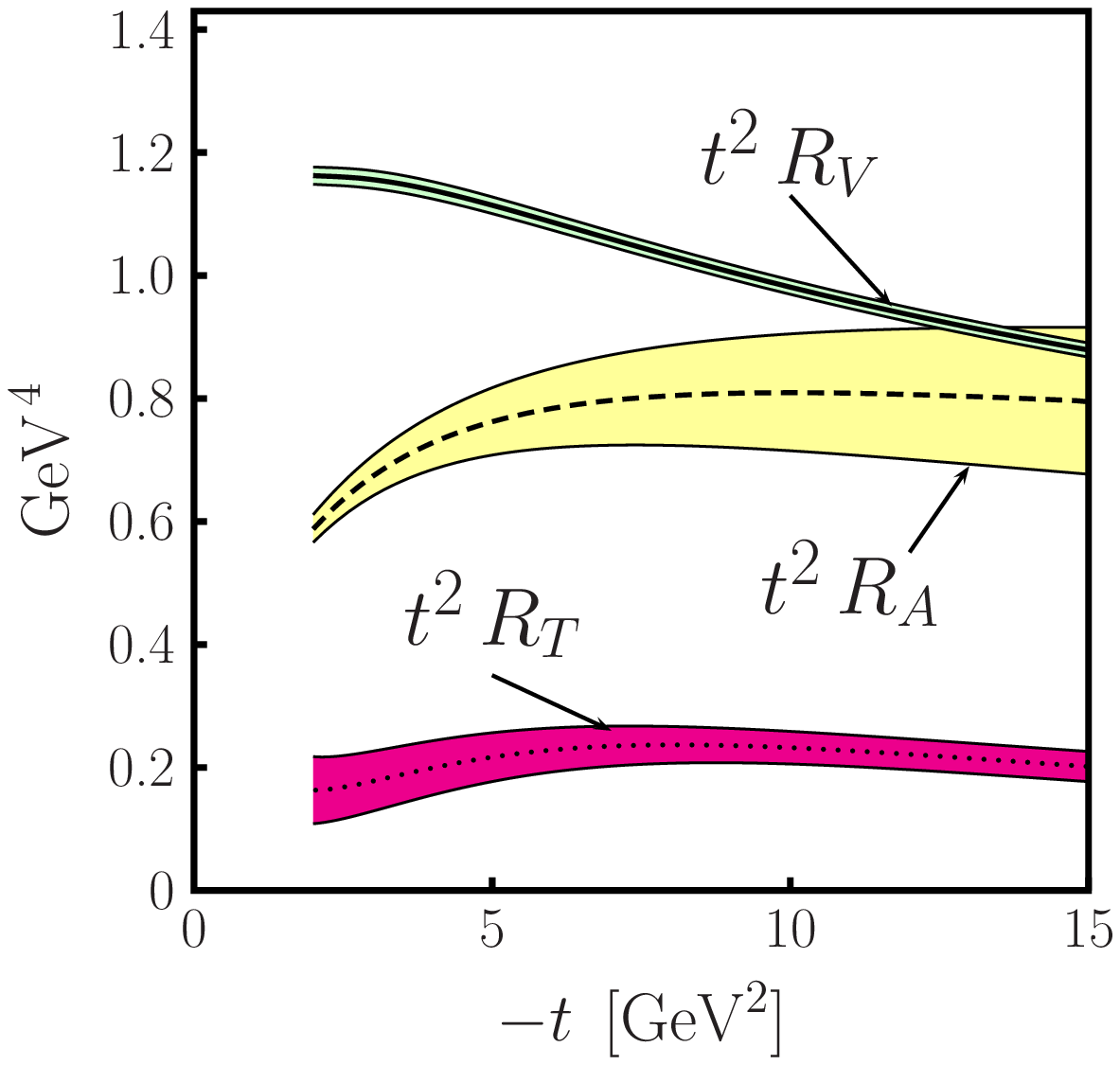}
\hspace*{2em}
\includegraphics[width=.47\textwidth, bb= 101 366 449 707,clip=true]
{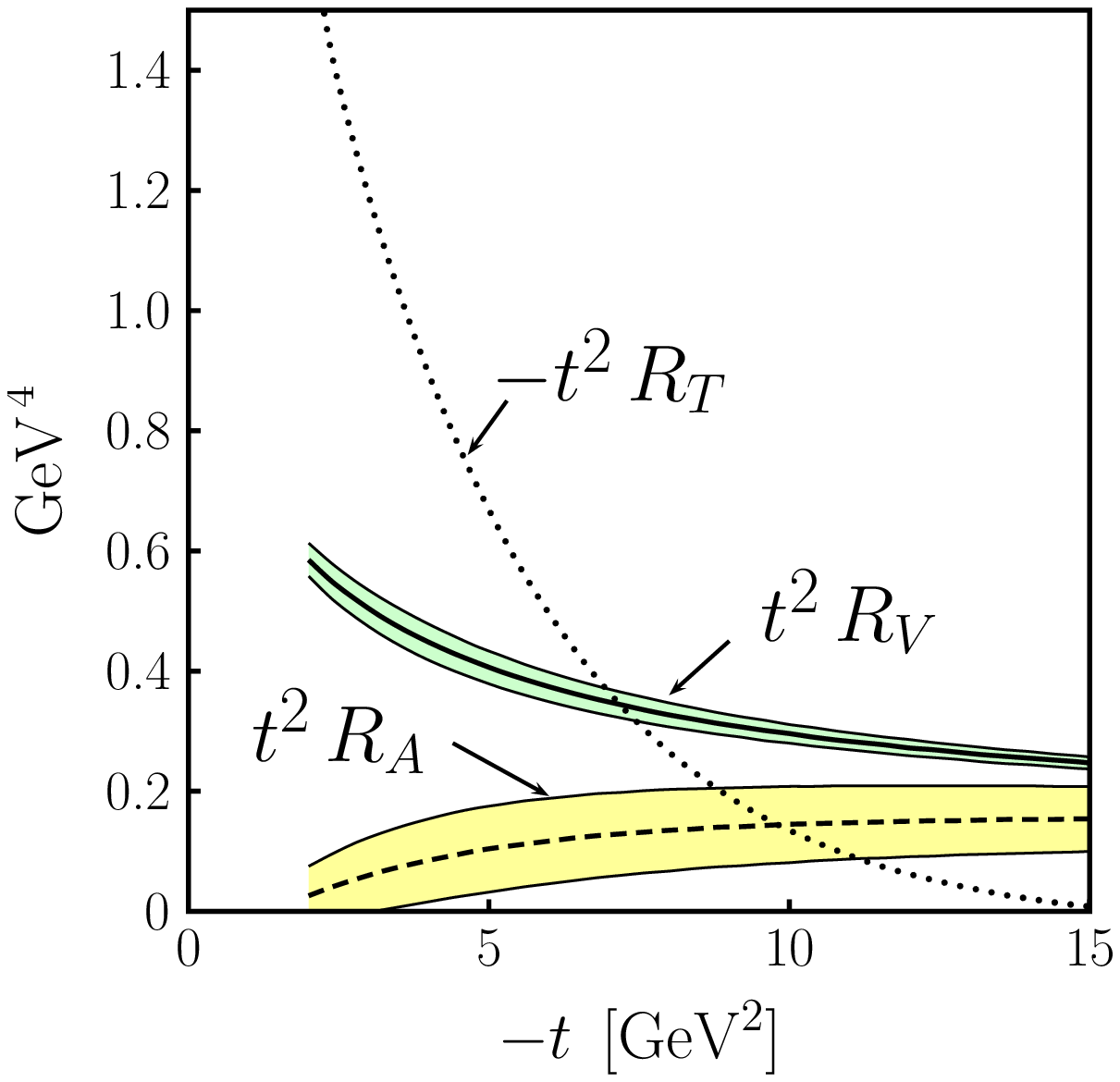}
\vspace*{-1.6em}
\caption{The Compton form factors for protons (left) and
  neutrons (right), scaled by $t^2$, evaluated from 
the GPDs determined in Ref.\ \protect\ci{DFJK4}. The bands represent
the parametric uncertainties of the form factors. For the tensor form
  factor of the neutron the error band is not shown for clearness.}
\label{fig:compton}
\end{figure}
\ba
R_V(t) &\simeq&\sum_{q=u,d} e_q^2 \int_{0}^1 \frac{dx}{x}\, H^q_v(x,t)\,, \nn\\
R_A(t) &\simeq& \sum_{q=u,d} e_q^2 \int_{0}^1 \frac{dx}{x}\, 
\widetilde{H}^q_v(x,t)\,, \nn\\[0.2em]
R_T(t) &\simeq&\sum_{q=u,d} e_q^2 \int_{0}^1 \frac{dx}{x}\, E^q_v(x,t)\,.
\label{Compton-formfactors}
\ea
In these expressions contributions from sea quarks have been 
neglected~\footnote{
An estimate of the sea quark contributions may be obtained by using 
the ansatz \req{ansatz} with the same profile function for the sea 
quarks as for the valence ones but replacing the valence quark density
with the CTEQ \ci{CTEQ} antiquark ones. The so estimated contributions 
are indeed small.}.
An analogue pseudoscalar form factor related to the GPD $\widetilde{E}$,
decouples in the symmetric frame. Numerical results for the Compton
form factors off protons and neutrons are shown in Fig.\
\ref{fig:compton}. The latter are obtained from Eq.\
\req{Compton-formfactors} with the help of isospin invariance:
$H^{d(u)}$ for the neutron equals $H^{u(d)}$ for the proton (see Eq.\
\req{pn-sr}). The axial-vector and the tensor form factors behave 
differently for protons and neutrons. This is a consequence of the
opposite signs for the $u$ and $d$ quark contributions (see Fig.\
\ref{fig:moment}) in combination with the different charge weigths. 
For smaller values of $-t$ the form factor $R_A$ cancels almost
completely in the case of the neutron. Since, at intermediate $-t$, 
$|E_v^d|$ exhibits a maximum that is more pronounced and located at 
smaller values of $x$ than that of $E_v^u$ (see Fig.\ \ref{fig:moment}), 
its $1/x$ moment is larger
than that of the latter one. In combination with the quark charges
this leads to due to a strong cancellation in the case of the proton
and, hence, to a very small form factor $R_T$. For the neutron, 
on the other hand, the properties of $E$ entaila a form factor $R_T$ 
which is negative and large in absolute value at intermediate $-t$. 

The handbag contribution leads to the following leading-order 
result for the Compton cross section \ci{DFJK1,HKM} 
\ba
\frac{d\sigma}{dt} &=& \frac{d\hat{\sigma}}{dt} \left\{ \frac12\, \Big[
R_V^2(t)\, + \frac{-t}{4m_p^2} R_T^2(t) + R_A^2(t)\Big] \right.\nn\\
&&\hspace*{-0.5cm}\left.  - \frac{u s}{s^2+u^2}\,
\Big[R_V^2(t)\,+ \frac{-t}{4m_p^2} R_T^2(t) - R_A^2(t)\Big]\right\}\,,
\label{dsdt}
\ea
where $d\hat{\sigma}/dt$ is the Klein-Nishina cross section for
Compton scattering off massless, point-like spin-1/2 particles of
charge unity. Next-to-leading order QCD corrections to the subprocess
have been calculated in Ref.\ \ci{HKM}. They are not displayed in 
\req{dsdt} but taken into account in the numerical results presented
below. 

\begin{figure}[t]
\includegraphics[width=.45\textwidth, bb= 149 289 569 715,clip=true]
{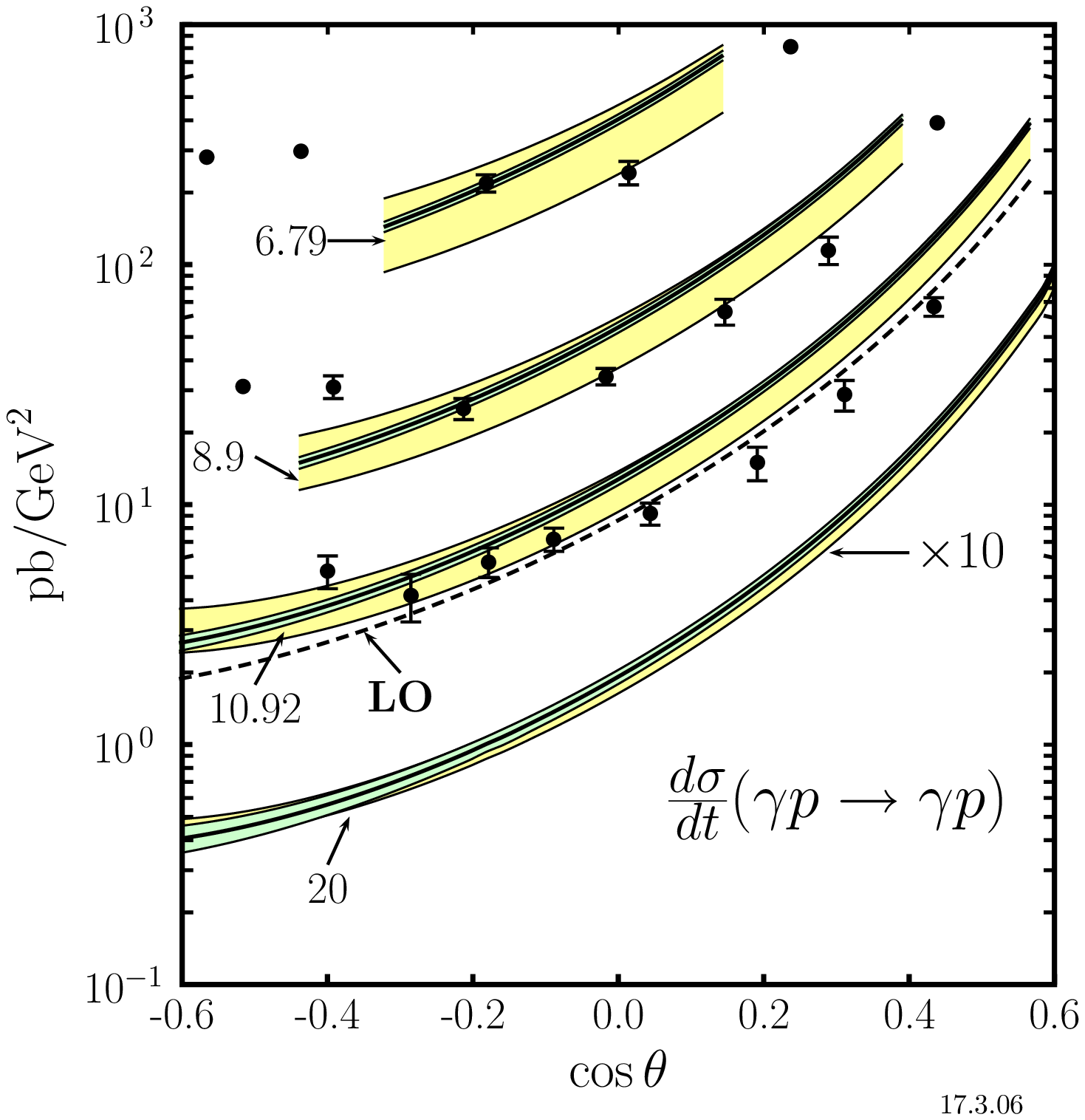}
\hspace*{2em}
\includegraphics[width=.45\textwidth,
  bb= 149 289 570 715,clip=true]{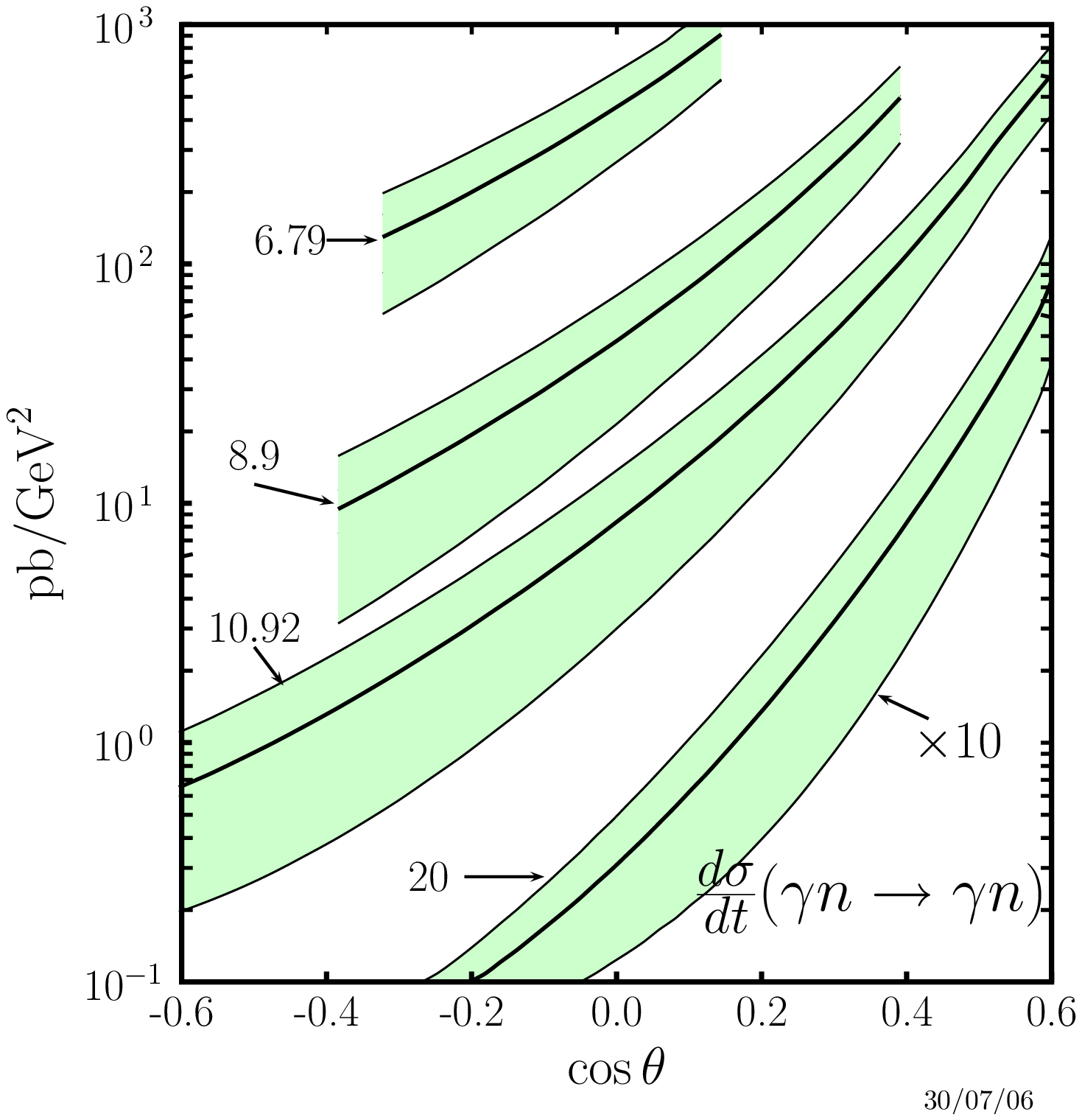}
\vspace*{-0.6em}
\caption{The cross sections for Compton scattering off protons (left)
  and neutrons (right) versus the c.m.s.\ scattering angle $\theta$ at 
  $s=6.9,\,8.9,\,11$ and $20\,\gev^2$. The error bands are explained
 in the text. Data are taken from Ref.\ \ci{nathan}.}
\label{fig:compton-cross}
\end{figure}

Inserting the Compton form factors \req{Compton-formfactors} into 
Eqs.\ \req{dsdt}, one can predict the Compton cross section in the 
wide-angle region. The results for sample values of $s$ are shown 
in Fig.\ \ref{fig:compton-cross} and compared to recent measurements from
JLab~\ci{nathan}. The inner bands of the predictions for $d\sigma/dt$
reflect the parametric errors of the form factors, essentially that 
of the vector form factor which dominates the cross section. 
The outer bands indicate the size of target mass corrections, see Ref.\
\cite{DFHK}. In order to comply with the kinematical requirements
for handbag factorization, at least in a minimal fashion, predictions 
are only shown for $-t$ and $-u$ larger than about $2.4\, \gev^2$. 
Fair agreement between theory and experiment is to be seen.
Predictions for Compton scattering off neutrons are also shown
in Fig.\ \ref{fig:compton-cross}. In the case of the neutron the
uncertainties of the form factors dominate. In the forward hemisphere
the cross section is large, comparable with the proton cross section
because the tensor form factors is so large, see Fig.\
\ref{fig:compton}. In the backward hemisphere, with increasing $-t$
the neutron cross section becomes small, tending towards the
suppression factor $(e_d/e_u)^4=1/16$ which limit is obvious in the
light of the discussion below Eq.\ \req{power}. A measurement of
$\gamma n \to \gamma n$ will provide a severe test of our present
understanding of the GPD $E$.
 
With the Compton form factors at hand also other Compton observables can
be calculated in a parameter-free way. Thus, for instance, the
helicity correlation, $A_{LL}$, between the initial state photon and
nucleon or, equivalently, the helicity transfer, $K_{LL}$, from the
incoming photon to the outgoing nucleon. In the handbag approach one
obtains \ci{HKM}
\be
A_{LL}\=K_{LL}\= \frac{s^2 - u^2}{s^2 + u^2}\, 
                    \frac{R_A(t)}{R_V(t)}\, \frac{1+\beta
                  \sqrt{-t}/(2m_p)R_T/R_V}{1-t/(4m^2_p)R_T^2/R_V^2}\,
                   \Big[1+\frac{R_A^2-R_V^2(1-t/(4m^2_p) R_T^2/R_V^2)}
                       {2R_V^2(1-t/(4m^2_p)R_T^2/R_V^2)}\,
                     \frac{t^2}{s^2+u^2}\Big]^{-1}\,,
\label{all}
\ee
where the factor in front of the form factors is the corresponding
observable for $\gamma q\to \gamma q$ ($\hat{A}_{LL}$). The kinematical 
factor $\beta$ reads $2m_p/\sqrt{s} \sqrt{-t}/(\sqrt{s}+\sqrt{-u})$. 
For Compton scattering off the proton Eq.\ \req{all} can be
approximated by
\be
A_{LL}\=K_{LL}\simeq \frac{s^2 - u^2}{s^2 + u^2}\,
\frac{R_A(t)}{R_V(t)}\,,
\label{all-proton}
\ee
The latter result is a
robust prediction of the handbag mechanism, the magnitude of the
subprocess helicity correlation, $\hat{A}_{LL}$, is only diluted
somewhat by the ratio of the form factors $R_A$ and $R_V$. It is to be
stressed that $A_{LL}$ and $K_{LL}$ are identically in the handbag
approach because the quarks are assumed to be massless and consequently 
there is no quark helicity flip. For an alternative approach, see 
Ref.\ \cite{miller}. For Compton scattering off neutrons $A_{LL}$ is
very small at JLab energies since $R_T$, neglected in Eq.\ \req{all-proton}, 
is large in this case.

The JLab Hall A collaboration \cite{hamilton} has presented a first
measurement of $K_{LL}$ at $s=6.9\,\gev^2$ and $t=-4\,\gev^2$.
The kinematical requirement of the handbag mechanism, $s,\; -t,\; -u
\gg m^2$, is not satisfied for this measurement since $-u$ is only
$1.13\,\gev^2$. One therefore has to be very cautious when comparing
this experimental result with the handbag predictions, there might be 
large dynamical and kinematical corrections. Nevertheless the
agreement of this data point with the prediction from the handbag
approach is good which may be considered as a non-trivial and
promising fact. Polarization data at higher energies are desired as
well as a measurement of the angle dependence.
\begin{figure}[t]
\includegraphics[width=0.39\textwidth,bb=119 340 434 685,clip=true]{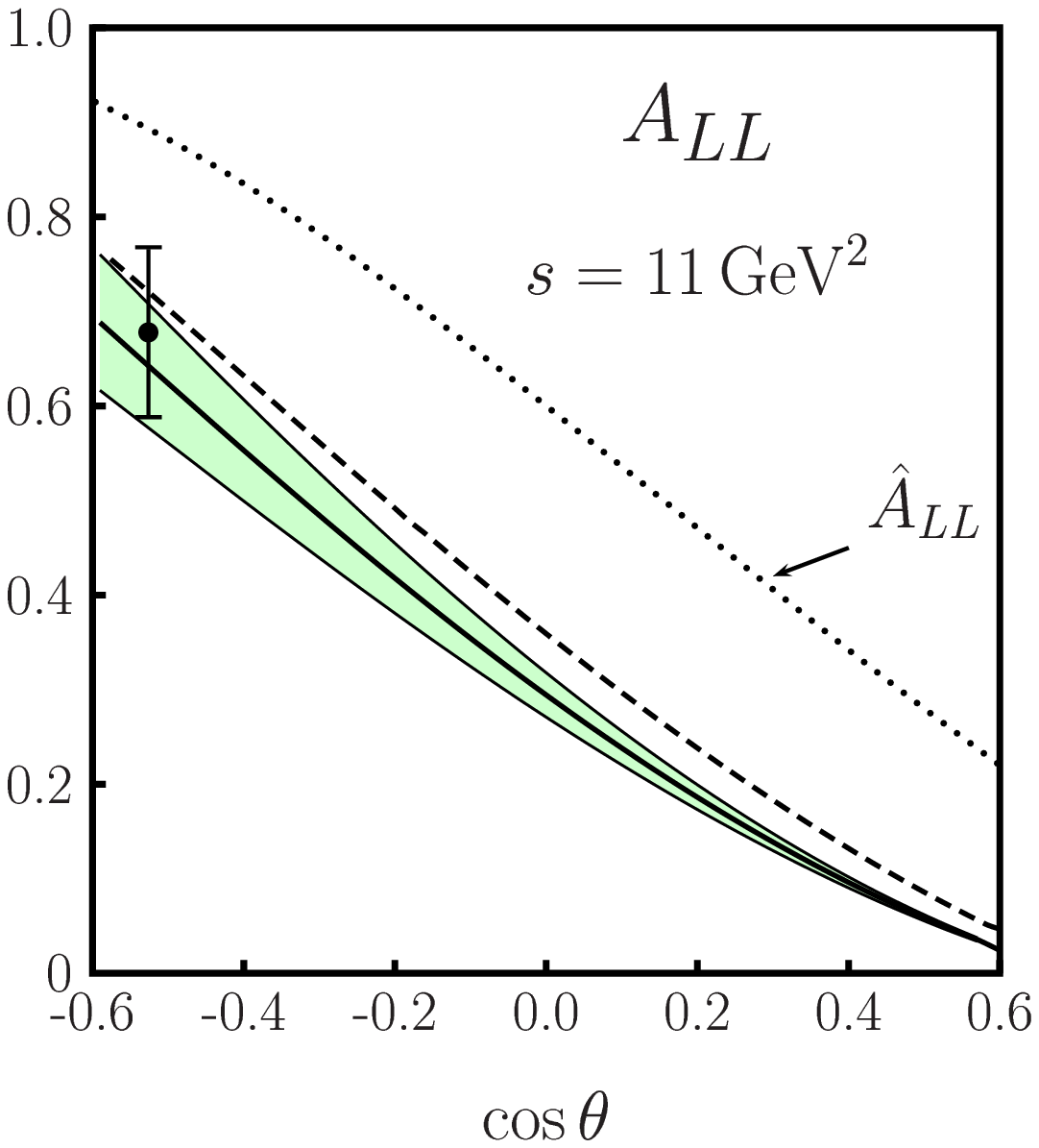}
\caption{Handbag prediction for the helicity correlation $A_{LL}=K_{LL}$ 
in Compton scattering off protons at $s=11\,\gev^2$ (right). Data are
taken from Ref.\ \ci{hamilton}. The green band represents the
uncertainties of the handbag prediction due to the errors of the
Compton form factors.}
\label{fig:all}
\end{figure}

The handbag approach also applies to wide-angle photo- and
electroproduction of pseudoscalar and vector mesons~\ci{hanwen}. The 
amplitudes again factorize into a parton-level subprocess, $\gamma
q\to M q$ now, and form factors which represent $1/x$-moments of
GPDs. Their flavor decomposition differs from those appearing in
Compton scattering. It now reflects the valence quark content of the 
produced meson. Since the GPDs and, hence, the form factors for a 
given flavor, $R_i^q$, $i=V,A,T$ are process independent they are
known from the analysis of Ref.~\cite{DFJK4} for $u$ and $d$ quarks. 
Therefore, the form factors occuring in photo-and electroproduction 
of pions and $\rho$ mesons within the handbag approach can also be
evaluated from the GPDs given in Ref.\ \ci{DFJK4}.

\section{Summary and outlook}
Results from a first analysis of the GPDs at zero skewness have been
reviewed. The analysis, performed in analogy to those of the usual
parton distributions, bases on a physically motivated parameterization
of the GPDs with a few free parameters fitted to the available nucleon 
form factor data. The analysis provides results on the valence-quark
GPDs $H$, $\widetilde{H}$ and $E$. The distribution of the quarks in
the impact parameter plane transverse to the direction of the nucleon's 
momentum can be evaluated from them. Polarizing the nucleon induces 
a flavor segregation in the direction orthogonal to the those of the 
nucleon's momentum and of its polarization. The average orbital angular 
momentum of the valence quarks can be estimated from the obtained
GPDs, too. 

The zero-skewness GPDs form the soft physics input to hard wide-angle 
exclusive reactions. For Compton scattering, for instance, the soft 
physics is encoded in specific form factors which represent $1/x$
moments of zero-skewness GPDs. Using the GPD determined in Ref.\
\ci{DFJK4}, one can evaluate the form factors and predict the
wide-angle Compton cross section. The results are found to be in good 
agreement with experiment.

An analysis as that one performed in Ref.\ \ci{DFJK4} can only be
considered as a first attempt towards a determination of the GPDs. It
needs improvements in various aspects. High quality data on the form
factors at larger $t$ are required in order to stabilize the
parameters. JLab will provide such data in the near future. As already 
mentioned CLAS \ci{CLAS} will come up with data on $G_M^n$ up to about 
$5\,\gev^2$. $G_E^n$ will be measured up to about $4\,\gev^2$ next
year  and $G_E^p$ up to $9\,\gev^2$ in 2007. 
The upgraded Jlab will allow measurements of the nucleon form factors
up to about $13\,\gev^2$. Data on the axial form factor are needed in
order to improve our knowledge of $\widetilde{H}$.

Up to now only the sum rules \req{pn-sr}, \req{axial-sr} have been
utilized in the GPD analysis. This, as discussed above, does not lead
to a unique solution for the GPDs. Higher moments are needed in order to lessen 
the dependence on the chosen parameterization for the GPDs. Such
moments can be provided by lattice gauge theories. The present lattice
results \ci{SESAM,qcdsf} are however calculated in scenarios where the
pion is heavy (typically $600 - 800\,\mev$) and the extrapolation
to the chiral limit is not performed. Obviously such
results are inappropriate for a GPD analysis. In a few years
however the quality of the moments from lattice gauge theories may
suffice. It is also tempting to use the data on wide-angle Compton
scattering~\ci{nathan} which provide information on $1/x$ moments of 
the GPDs. Even at the highest measured energy, $s=11\,\gev^2$, the 
Compton cross section may still be contaminated by power corrections 
rendering its use in a GPD analysis dubious. For data at, say, 
$s=20\,\gev^2$ the situation may be different. 

Up to now only the zero-skewness GPDs have been determined. Although
the sum rules \req{pn-sr}, \req{axial-sr} holds for all values of the 
skewness it seems hopeless task to extract the GPDs as funtions of three
variables from them. Additional information is demanded in the case on
$\xi \neq 0$ and will be provided by deeply virtual exclusive
scattering (DVES). 
\section*{Acknowledgements} It is a pleasure to thank Aaron Bernstein
and Costas Papanicolas for organising this interesting
meeting and for the hospitality extended to the author in Athens.


\begin{thebibliography}{99}
 
\bibitem{CTEQ} J.\ Pumplin {\it et al} [CTEQ collaboration], 
JHEP {\bf 0207}, 012 (2002). 

\bibitem{happex} F.~E.~Maas {\it et al.},
  Phys.\ Rev.\ Lett.\  {\bf 94}, 152001 (2005); 
K.~A.~Aniol {\it et al.}  [HAPPEX Collaboration],
  nucl-ex/0506011;
D.~S.~Armstrong {\it et al.}  [G0 Collaboration],
  Phys.\ Rev.\ Lett.\  {\bf 95}, 092001 (2005). 

\bibitem{HERMES} A.~Airapetian {\it et al.}  [HERMES Collaboration],
Phys.\ Rev.\ Lett.\ {\bf 92}, 012005 (2004).

\bibitem{BB}J.~Bl{\"u}mlein and H.~B{\"o}ttcher,
Nucl.\ Phys.\ B{\bf 636}, 225 (2002).

\bibitem{DFJK4} M.~Diehl, T.~Feldmann, R.~Jakob and P.~Kroll,
Eur.\ Phys.\ J.\ C {\bf 39}, 1 (2005). 

\bibitem{guidal} M.~Guidal, M.~V.~Polyakov, A.~V.~Radyushkin and M.~Vanderhaeghen,
Phys.\ Rev.\ D {\bf 72}, 054013 (2005). 

\bibitem{burk} M.~Burkardt,
Int.\ J.\ Mod.\ Phys.\ A {\bf 18}, 173 (2003).

\bibitem{DFJK1} M.~Diehl, T.~Feldmann, R.~Jakob and P.~Kroll,
Eur.\ Phys.\ J.\ C {\bf 8}, 409 (1999).

\bibitem{rad98} A.~V.~Radyushkin,
Phys.\ Rev.\ D {\bf 58}, 114008 (1998).

\bibitem{arbarbanel} H.D.I.\ Arbarbanel, M.L.\ Goldberger and S.B.\
  Treiman, Phys.\ Rev.\ Lett.\ {\bf22}, 500 (1969);
P.V.\ Landshoff, J.C.\ Polkinghorne and R.D.\ Short,
Nucl.\ Phys.\ B {\bf 28}, 225 (1971).

\bibitem{DFJK3} M.~Diehl, T.~Feldmann, R.~Jakob, and P.~Kroll,
Nucl.\ Phys.\ B {\bf 596}, 33 (2001),
Erratum-ibid.\ {\bf 605}, 647 (2001).

\bibitem{DY} S.~D.~Drell and T.~M.~Yan,
Phys.\ Rev.\ Lett.\ {\bf 24}, 181 (1970).

\bibitem{CLAS} W.~K.~Brooks and J.~D.~Lachniet,
  Nucl.\ Phys.\ A {\bf 755}, 261 (2005).

\bibitem{ji97} X.-D. Ji,
Phys.\ Rev.\ Lett.\ {\bf 78}, 610 (1997).

\bibitem{SESAM} P.~H\"agler, J.~W.~Negele, D.~B.~Renner, W.~Schroers,
  T.~Lippert and K.~Schilling [LHPC Collaboration],
Eur.\ Phys.\ J.\ A {\bf 24S1}, 29 (2005). 

\bibitem{qcdsf}M.~Gockeler, R.~Horsley, D.~Pleiter, P.~E.~L.~Rakow,
  A.~Schafer, G.~Schierholz and W.~Schroers [QCDSF Collaboration],
  Phys.\ Rev.\ Lett.\  {\bf 92}, 042002 (2004).

\bibitem{sivers}D.W.\ Sivers, Phys.\ Rev.\ D {\bf 43}, 261 (1991).

\bibitem{burkardt06} M.~Burkardt,
  Phys.\ Rev.\ D {\bf 72}, 094020 (2005).

\bibitem{HERMES05} HERMES collaboration, Phys.\ Rev.\ Lett.\ {\bf 94},
  012002 (2005).

\bibitem{HKM} H.~W.~Huang, P.~Kroll and T.~Morii,
Eur.\ Phys.\ J.\ C {\bf 23}, 301 (2002)[Erratum-ibid.\ C {\bf 31}, 279 (2003)]. 

\bibitem{nathan} A.~Danagulian, A.~M.~Nathan, M.~Roedelbronn,
  D.~J.~Hamilton, C.~E.~Hyde-Wright, V.~H.~Mamian and
  B.~Wojtsekhowski, 
Nucl.\ Phys.\ A {\bf 755}, 281 (2005);
V.~H.~Mamian {\it et al.}, proceedings of the Armenyan Academy of
Science, Physics, {\bf 40}, 325 (2005).

\bibitem{DFHK} M.~Diehl, T.~Feldmann, H.~W.~Huang and P.~Kroll,
Phys.\ Rev.\ D {\bf 67}, 037502 (2003). 

\bibitem{miller} G.~A.~Miller,
  Phys.\ Rev.\ C {\bf 69}, 052201 (2004).

\bibitem{hamilton} D.~J.~Hamilton {\it et al.}  [Jefferson Lab Hall A Collaboration],
  Phys.\ Rev.\ Lett.\  {\bf 94}, 242001 (2005).

\bibitem{hanwen} H.~W.~Huang and P.~Kroll,
Eur.\ Phys.\ J.\ C {\bf 17}, 423 (2000);
H.~W.~Huang, R.~Jakob, P.~Kroll and K.~Passek-Kumericki,
Eur.\ Phys.\ J.\ C {\bf 33}, 91 (2004).

\end{thebibliography}
\end{document}